\documentclass[12pt,3p,sort&compress,preprint]{elsarticle}

\usepackage{graphicx}
\usepackage{bm} 
\usepackage{epsfig}
\usepackage[latin1]{inputenc}
\usepackage{float,amsmath}
\usepackage{amssymb}

\def\lsim{\mathrel{\rlap{\lower4pt\hbox{$\sim$}}
    \raise1pt\hbox{$<$}}}                
\def\gsim{\mathrel{\rlap{\lower4pt\hbox{$\sim$}}
    \raise1pt\hbox{$>$}}}                
\newcommand{\beq}{\begin{equation}}
\newcommand{\eeq}{\end{equation}}
\newcommand{\bqa}{\begin{eqnarray}}
\newcommand{\eqa}{\end{eqnarray}}
\newcommand{\phard}{p_{\rm hard}}
\newcommand{\nn}{\nonumber}

\usepackage[usenames]{color}

\begin{document}


\begin{frontmatter}

\title{Non-boost-invariant anisotropic dynamics}

\author[itp,fias]{Mauricio Martinez}
\address[itp]{
Institut f\"{u}r Theoretische Physik \\
Goethe-Universit\"{a}t Frankfurt \\
Max-von-Laue Strasse 1\\
D-60438, Frankfurt am Main, Germany
}
\author[gettysburg,fias]{Michael Strickland}
\address[gettysburg]{
  Physics Department, Gettysburg College\\
  Gettysburg, PA 17325 United States
}
\address[fias]{
Frankfurt Institute for Advanced Studies\\
Ruth-Moufang-Strasse 1\\
D-60438, Frankfurt am Main, Germany
}

\begin{abstract}
We study the non-boost-invariant evolution of a quark-gluon plasma subject to large early-time momentum-space anisotropies. 
Rather than using the canonical hydrodynamical expansion of the distribution function around an isotropic equilibrium state, we 
expand around a state which is anisotropic in momentum space
and parameterize this state in terms of three proper-time and spatial-rapidity dependent parameters.  
Deviations from the Bjorken scaling solutions are naturally taken into account by the time evolution of the spatial-rapidity dependence
of the anisotropic ansatz. As a result, we obtain three coupled partial differential equations 
for the momentum-space anisotropy, the typical momentum of the degrees of freedom, and the longitudinal flow. Within this 
framework (0+1)-dimensional Bjorken expansion is obtained as an asymptotic limit. Finally, we make quantitative comparisons of the 
temporal and spatial-rapidity evolution of the dynamical parameters and resulting
pressure anisotropy in both the strong and weak coupling limits. 
\end{abstract}



\end{frontmatter}


\section{Introduction}
\label{sec:introduction}

In recent years relativistic viscous hydrodynamical models have been applied with great success to the description of the non-central 
anisotropic flow measured at the Relativistic Heavy Ion Collider
~\cite{Huovinen:2001cy, Hirano:2002ds,Tannenbaum:2006ch, Kolb:2003dz,Dusling:2007gi,Luzum:2008cw,Song:2008hj,Heinz:2009xj}. 
Despite this success, it is well known 
viscous hydrodynamics is not an accurate description during early times after the initial impact. The hot and 
dense matter created after the collision is rather small in transverse extent and expands very 
rapidly in the longitudinal direction. In addition, transverse expansion of the matter is not expected to become important until
times on the order of 2 or 3 fm/c after the collision.  
As a consequence, large momentum-space anisotropies are developed at early times. When these anisotropies are 
large the shear viscous corrections can be of the same order or larger than the isotropic pressure, casting doubt on the validity of 
viscous hydrodynamics to faithfully model the dynamics. For example, it is known that within the framework of 2nd-order
viscous hydrodynamics large momentum-space anisotropies can lead to negative longitudinal pressure at early 
times~\cite{Martinez:2009mf}.

One is therefore motivated to develop an alternative approach that will allow us to better understand the transition between 
early-time anisotropic expansion and late-time hydrodynamical behaviour in ultrarelativistic heavy ion collisions.  The work 
presented here fills this gap using a technique which allows one to smoothly connect both regimes using a unified theoretical
framework. We present a method which allows us to derive hydrodynamical-like evolution equations in the presence of large 
momentum-space anisotropies. The method consists of changing the usual hydrodynamic expansion of the one-particle distribution 
function in the local rest frame
\beq
\label{eq:oldexp}
f({\bf x},{\bf p},\tau) = f_{\rm eq}(|{\bf p}|,T(\tau)) + \delta f_1 + \delta f_2 + \cdots \; ,
\eeq
which is an expansion around an isotropic equilibrium state $f_{\rm eq}(|{\bf p}|,T)$, to one
in which the expansion point itself can contain momentum-space anisotropies
\beq
f({\bf x},{\bf p},\tau) = f_{\rm aniso}({\bf p},\phard,\xi) + \delta f_1^\prime + \delta f_2^\prime + \cdots \; .
\label{eq:newexp}
\eeq
In Eq.~(\ref{eq:newexp}) $\xi$ is a parameter that measures the amount of momentum-space anisotropy and
$\phard$ is a non-equilibrium momentum scale which can be identified with the temperature of the system only in the limit of isotropic
equilibrium. We introduce an ansatz for  the leading order term of Eq.~(\ref{eq:newexp}), $f_{\rm aniso}({\bf p},\phard,\xi)$, which 
approximates the system via spheroidal equal occupation number surfaces as opposed to the spherical ones associated with the isotropic 
equilibrium distribution function~\cite{Romatschke:2003ms}. We do not calculate the remaining terms of the corrections 
$\delta f_n^\prime$, their evaluation is beyond the scope of this paper. Nevertheless, it may be possible to perform such a calculation 
in a consistent way for relativistic system with the use of irreducible anisotropic tensors~\cite{kroger} or spheroidal harmonics
~\cite{abramowitz_stegun64} for which our ansatz is the leading order term. Moreover, one expects that the corrections 
$\delta f_n^\prime$ will have smaller magnitude than the isotropic corrections $\delta f_n$ because of the 
momentum-space anisotropy is built into the leading order term in the reorganized expansion. 

In this paper we extend our previous work on highly anisotropic dynamics~\cite{Martinez:2010sc} 
to include both temporal and spatial-rapidity dependence of the system thereby removing the assumption of boost invariance of the system.
By making use of an ansatz for $f_{\rm aniso}({\bf p},\phard,\xi)$, we derive leading order evolution equations for a non-boost-invariant plasma which expands 
only in the longitudinal direction.  Based on this ansatz we obtain three coupled partial differential equations for the momentum-space anisotropy, the
typical momentum scale of the particles and the longitudinal flow rapidity.  We show that these equations reduce to the boost-invariant
description of the plasma in the appropriate limit.  Finally, we present numerical solutions of the non-boost-invariant 
evolution equations and make a quantitative description of the pressure anisotropy for weak and strong coupling. 
 

\section{Kinetic theory approach for a non-boost-invariant and highly anisotropic QGP}
\label{sec:setup}

In this section we derive the equations of motion necessary to describe the transition between early time dynamics and 
late-time hydrodynamical behaviour of a non-boost-invariant system. For simplicity, we assume that there is no transverse 
expansion and that all expansion is 
along the longitudinal direction.\footnote{We point out that the method presented 
here can be generalized to include azimuthal and non-azimuthal transverse expansion. 
We postpone these studies to a future publication~\cite{Mart-Strick}.} The derivation is based on taking moments of the 
Boltzmann equation. Before presenting the evolution equations, we first review the moment method of the Boltzmann equation applied 
to general relativistic theories~\cite{GLW,Israel:1979wp,Muronga:2006zx}, then we apply this formalism to the case of a 
non-boost-invariant system which is anisotropic in momentum space. 


\subsection{Moments of the Boltzmann equation}
\label{subsec:moment}

For a dilute system, the Boltzmann relativistic transport equation for the on-shell phase space density $f({\bf x},{\bf p},t)$ 
is~\cite{GLW}
\beq
\label{eq:Boltzmanneq}
p^\mu\partial_\mu f({\bf x},{\bf p},t)=-{\cal C}\bigl[f\bigr]\,,
\eeq
where ${\cal C}\bigl[f\bigr]$ is the collisional kernel which is a functional of the distribution function. In order to have a 
tractable approach to dissipative dynamics from kinetic theory, one usually takes 
moments of the Boltzmann equation~(\ref{eq:Boltzmanneq}) to determine the equations of motion of the dissipative 
currents~\cite{GLW,Israel:1979wp,Muronga:2006zx}.

The equation of motion for the n-th moment of the Boltzmann equation is\footnote{Our notation for the  
momentum-space integral is $\int_{\bf p}\equiv (2\pi)^{-3}\int d^3{\bf p}/p^0$.}
\beq
\label{eq:generalmoment}
\partial_{\mu_{n+1}} I^{\mu_1\mu_2\ldots\mu_{n+1}}= -P^{\mu_1\mu_2\ldots\mu_{n}}\,,
\eeq
where 
\begin{subequations}
 \begin{align}
  I^{\mu_1\mu_2\ldots\mu_{n+1}}&=\int_{\bf p} p^{\mu_1}p^{\mu_2}\ldots p^{\mu_{n+1}}\,f({\bf x},{\bf p},t)\,,
\label{eq:Igeneral}\\
  P^{\mu_1\mu_2\ldots\mu_{n}}&=\int_{\bf p} p^{\mu_1}p^{\mu_2}\ldots p^{\mu_{n}}\,{\cal C}\bigl[f\bigr]\,.
\label{eq:Pgeneral}
 \end{align}
\end{subequations}
Eq.~(\ref{eq:generalmoment}) defines an infinite set of coupled equations for these moments. By truncating the expansion, one obtains a finite set of equations~\cite{GLW}. 
In the Israel-Stewart (IS) theory, the evolution equation for the dissipative currents  and the transport coefficients can be extracted from the second moment of the Boltzmann equation within the 14 Grad's ansatz for the distribution function~\cite{Israel:1979wp,Muronga:2006zx}. 

By using Eq.~(\ref{eq:generalmoment}) for $n=0, 1, 2$; one finds
\begin{subequations}
\label{eq:dissipativeeqs}
\begin{align}
\int_{\bf p} p^\mu \partial_\mu f(x,p) &\equiv \partial_\mu N^\mu = - \int_{\bf p} {\cal C} [f] ~,\label{eq:Number}\\
\int_{\bf p}  p^\mu p^\nu \partial_\mu f(x,p) &\equiv \partial_\mu T^{\mu\nu} =
- \int_{\bf p}  p^\nu{\cal C} [f] \equiv 0 ~,\label{eq:ene-mom}\\
\int_{\bf p} p^\mu p^\nu p^\lambda \partial_\mu f(x,p) &\equiv \partial_\mu I^{\mu\nu\lambda} =
- \int_{\bf p} p^\nu p^\lambda {\cal C} [f] \equiv - P^{\nu\lambda}~. \label{eq:Fluxes}
\end{align}
\end{subequations}
The zeroth moment of the Boltzmann equation~(\ref{eq:Number}), tells us the evolution of the particle current $N^\mu$. When 
$\int_{\bf p}{\cal C}[f]=0$, we have particle number conservation, however, for general gauge theories like QCD where inelastic 
processes are important, away from chemical equilibrium 
we have $\int_{\bf f}{\cal C}[f]\neq 0$, i.e. there is a source term for particle production and annihilation
~\cite{Biro:1993qt,Baier:2000sb,Xu:2007aa,Xu:2004mz}. The first moment of the Boltzmann equation~(\ref{eq:ene-mom}), gives us the 
conservation of the energy and momentum since $\int_{\bf p} p^\mu {\cal C}[f]=0$. Eq.~(\ref{eq:Fluxes}) represents the balance 
of the dissipative fluxes. $I^{\mu\nu\lambda}$ is a symmetric tensor of the dissipative fluxes and $P^{\nu\lambda}$ is a traceless 
tensor that is related to dissipative processes due to collisions~\cite{GLW, Muronga:2006zx,Romatschke:2009im}.  
The right hand side of Eq.~(\ref{eq:Fluxes}) does not trivially vanish, so in general 
$\int_{\bf p}p^\nu p^\lambda{\cal C}[f]\equiv P^{\nu\lambda}\neq 0$.

To obtain the complete set of equations for the dynamical variables such as the particle density $n$, energy density ${\cal E}$, etc., in hydrodynamical treatments one canonically expands the distribution function around equilibrium by using some functional form for the fluctuations (see Eq.~(\ref{eq:oldexp})). In our approach, we do not expand the distribution function around an isotropic expansion point.
Instead, we find evolution equations for the kinematical parameters contained in the leading term $f_{\rm aniso}$ in Eq.~(\ref{eq:newexp}) by taking moments of the Boltzmann equation.


\subsection{Energy-momentum tensor for a highly anisotropic plasma}
\label{sec:en-momtensanis}

The general energy-momentum tensor for an anisotropic plasma is given by~\cite{Florkowski:2010cf,Ryblewski:2010bs}
\beq
T^{\mu \nu} = \left( {\cal E}  + {\cal P}_T\right) u^{\mu}u^{\nu} 
-{\cal P}_T \, g^{\mu\nu} - ({\cal P}_T - {\cal P}_L) v^{\mu}v^{\nu}\,, 
\label{Tmunudec}
\eeq
where ${\cal E}$, ${\cal P}_T$, and ${\cal P}_L$ are the energy density, transverse pressure, and longitudinal pressure. $u^\mu$ is a 
unitary four-vector which defines the longitudinal flow velocity of the system. In the local rest frame (LRF), $u^\mu =(1,0,0,0)$. In addition to 
the fluid velocity, we introduce a vector $v^\mu$ that is a normalized space-like vector ($v^\mu v_\mu = -1$) which is orthogonal to 
$u^\mu$ ($u_\mu v^\mu =0$). In the LRF, $v^\mu =(0,0,0,1)$.\footnote{Usually in hydrodynamics, one uses operator 
$\Delta^{\mu\nu}=g^{\mu\nu}-u^\mu u^\nu$ which is orthogonal to the four-velocity $u^\mu$. From its definition, it is straightforward
to see that $v^\mu$ is an eigenvector of the operator $\Delta^{\mu\nu}$ with eigenvalue equal to one, i.e. $\Delta^{\mu\nu}v_\nu =v^\mu$.}

The energy momentum tensor takes a diagonal form $T^{\mu\nu}= \text{diag}({\cal E},{\cal P}_T,{\cal P}_T,{\cal P}_L)$ in the LRF. The evolution of its components are given by energy-momentum conservation 
\beq
\label{eq:conslaw}
\partial_\mu T^{\mu\nu}=0\,.
\eeq
If we project the last expression with $u^\mu$ and $v^\mu$ we obtain~\cite{Ryblewski:2010bs}
\begin{subequations}
\label{eq:en-mom-eqs}
\begin{align}
{\cal D} {\cal E}&= -\bigl({\cal E}+{\cal P}_T\bigr)\theta+\bigl({\cal P}_T - {\cal P}_L\bigr)u_\mu{\bar{\cal D}}v^\mu \,,\label{eq:ene}\\
{\bar{\cal D}}{\cal P}_L&=\bigl({\cal P}_T - {\cal P}_L\bigr)\bar{\theta}+\bigl({\cal E}+{\cal P}_T\bigr)v_\mu{\cal D}u^\mu\,,
\label{eq:longpresseq}
\end{align}
\end{subequations}
where ${\cal D}=u^\mu\partial_\mu $, ${\bar{\cal D}}=v^\mu\partial_\mu $, $\theta=\partial_\mu u^\mu$ and $\bar{\theta}=\partial_\mu v^\mu$. 


\subsection{Boltzmann equation for the anisotropic distribution function}
\label{subsec:RSansatz}

In order to make analytical progress, we will assume that the leading term of Eq.~(\ref{eq:newexp}) is given by the ansatz proposed by Romatschke and Strickland~\cite{Romatschke:2003ms}. Within this approach the leading order anisotropic distribution function in the LRF can be obtained from an arbitrary isotropic distribution  function ($f_{\rm iso}$) by squeezing or stretching
$f_{\rm iso}$ along one direction in momentum space
\beq
f({\bf x},{\bf p},\tau)= f_{\rm iso}\bigl(p_\mu p_\nu\xi^{\mu\nu}(x)/p^2_{\rm hard}(x)\bigr) \; ,
\label{eq:rsansatz}
\eeq
where $p_{\rm hard}$ is related to the average momentum of the partons and $\xi^{\mu\nu}(x)$ is a symmetric tensor
which indicates the direction and strength of the anisotropy. This particular choice is a generalization of 
the original longitudinal RS ansatz~\cite{Romatschke:2003ms}. 

In this work we restrict ourselves to a system which expands along the longitudinal direction in a non-boost-invariant 
way so that $\xi^{\mu\nu}$ takes the form
\beq
\label{eq:xinumu}
\xi^{\mu\nu}(x)= \text{diag} (1,0,0,\xi(x)),
\eeq
where $-1 < \xi < \infty$ is a parameter that reflects the strength and type of anisotropy.
\footnote{The anisotropy parameter $\xi$ is related to the average longitudinal and transverse momentum of the plasma partons via 
the relation~\cite{Mauricio:2007vz,Martinez:2008di} $\xi=\frac{1}{2} \langle p_T^2\rangle/\langle p_L^2\rangle - 1$.  
The system is locally isotropic when $\xi=0$. 
}
Using this, the first argument of Eq.~(\ref{eq:rsansatz}) becomes
\bqa
\xi^{\mu\nu}(x)\,p_\mu p_\nu &=& (p_{0})^2 +\xi p_z^2\,,\nonumber\\
&=& p_T^2 +(1+\xi)p_z^2\,,
\eqa
where we have considered on-shell massless particles and $p_T^2=p_x^2+p_y^2$. Choosing $\xi^{\mu\nu}$ as given by 
(\ref{eq:xinumu}) allows us to recover the original form of the RS ansatz as used by the authors of Ref.~\cite{Romatschke:2003ms}, i.e.
\beq
\label{eq:rsansatz-long}
f({\bf x},{\bf p},\tau)=f_{\rm RS}({\bf p},\xi,\phard)=f_{\rm iso}([{\bf p}_T^2+(1+\xi)p_z^2]/\phard^2) \, .
\eeq
To write the non-boost-invariant Boltzmann equation for the RS ansatz~(\ref{eq:rsansatz-long}), it is convenient to switch to
the coordinates specified by $\tau=\sqrt{t^2-z^2}$ and $\tanh \varsigma = z/t$. In this coordinate system, the RS ansatz~(\ref{eq:rsansatz-long}) is 
\beq
f_{\rm RS}({\bf p},\xi,\phard,\vartheta)=f_{\rm iso}\bigl({\bf p}_T^2[1+(1+\xi)\sinh^2(y-\vartheta)]/\phard^2\bigr)\,, 
\eeq
where $\vartheta$ is the hyperbolic angle associated with the comoving or LRF and $y$ is the lab momentum-space rapidity. Note that $\xi$, $\phard$ and $\vartheta$ are functions of both $\tau$ and $\varsigma$.
After changing variables, the Boltzmann equation is
\beq
\label{eq:Boltztaueta}
\left(p_T \cosh (y-\vartheta)\frac{\partial}{\partial\tau}+
\frac{p_T\,\sinh (y-\vartheta)}{\tau}\,\frac{\partial}{\partial\varsigma}\right) f_{\rm RS}({\bf p},\xi,\phard,\vartheta) = 
-{\cal C}[f_{\rm RS}]\,.
\eeq
So far we have not prescribed a specific collisional kernel ${\cal C}[f]$. Since the calculation of the collisional kernel is difficult in general, we model it here by considering the relaxation time approximation
\beq
{\cal C}[f_{RS}] = p_\mu u^\mu \, \Gamma \, \left[ f_{\rm RS}({\bf p},\xi,\phard,\vartheta) - f_{\rm eq}(|{\bf p}|,T) \right]\,,
\label{rel-time}
\eeq
where $f_{\rm eq}(|{\bf p}|,T)$ is the local equilibrium distribution and $\Gamma$ is the relaxation rate. 
The temperature $T$ is determined dynamically by requiring energy conservation 
${\cal E}_{\rm eq}(\tau)={\cal E}_{\rm non\hbox{-}eq}(\tau)$~\cite{Martinez:2010sc,Baym:1984np}. From the results of~\ref{Ap:A} one finds for the RS ansatz~(\ref{eq:rsansatz-long}) that
$T=R^{1/4}(\xi)\phard$~\cite{Martinez:2010sc,Martinez:2009ry}.  As we demonstrated in a previous paper \cite{Martinez:2010sc},
$\Gamma$ can be related to the shear viscosity and shear relaxation time of the system by matching our
treatment to 2nd-order viscous hydrodynamics in the small anisotropy limit.  We will return to this point
in the end of the next section.


\subsection{Evolution equations for the kinematical parameters of the RS ansatz}
\label{subsec:non-boost}

In this section we apply the moment method to the anisotropic RS ansatz~(\ref{eq:rsansatz-long}) for a non-boost-invariant system. 
The boost invariant case was already studied in a previous publication~\cite{Martinez:2010sc}.   We will assume in what
follows that the system is conformal and therefore ${\cal E}_{\rm iso} = 3 {\cal P}_{\rm iso}$.

For the longitudinal dimension, it is useful to use the following parametrizations for the vectors $u^\mu$ and $v^\mu$ introduced in Sect.~\ref{sec:en-momtensanis}~\cite{Ryblewski:2010bs,Satarov:2006jq,Satarov:2006iw}
\begin{subequations}
\begin{align}
u^\mu &= (\cosh \vartheta(\tau,\varsigma),0,0,\sinh \vartheta(\tau,\varsigma)) \, , \\
v^\mu &= (\sinh \vartheta(\tau,\varsigma),0,0,\cosh \vartheta(\tau,\varsigma)) \, ,
\end{align}
\label{eq:vectors}
\end{subequations}
where $\vartheta$ is the hyperbolic angle associated with the velocity of the LRF as measured in the lab frame. With this particular choice, one finds
\begin{eqnarray}
{\cal D} &=&  \cosh (\vartheta-\varsigma)\,\partial_\tau+\frac{\sinh (\vartheta-\varsigma)}{\tau}\,\partial_\varsigma \,,\\
\bar{\cal D} &=&   \sinh (\vartheta-\varsigma)\,\partial_\tau +\frac{\cosh (\vartheta-\varsigma)}{\tau}\,\partial_\varsigma \,.
\end{eqnarray}
and
\begin{eqnarray}
&& \theta= u_\mu {\bar{\cal D}}v^\mu =  \Bigl(\sinh (\vartheta-\varsigma)\,\partial_\tau +\frac{\cosh (\vartheta-\varsigma)}{\tau}\,\partial_\varsigma \Bigr)\vartheta\,,\\
&& \bar{\theta} = -v_\mu {\cal D}u^\mu = \Bigl(\cosh (\vartheta-\varsigma)\,\partial_\tau+\frac{\sinh (\vartheta-\varsigma)}{\tau}\,\partial_\varsigma \Bigr) \vartheta\,.
\label{eq:thetas}
\end{eqnarray}

\subsubsection{0th moment of the non-boost invariant Boltzmann equation}
\label{subsubsec:0th-Bolteqn}

The 0th moment of the Boltzmann equation is equivalent to the equation for the 
evolution of the particle current $N^\mu = n u^\mu$ (Eq.~(\ref{eq:Number})) which can 
be written compactly as
\beq
{\cal D} n + n \theta = -\int_{\bf p} {\cal C}[f] \, ,
\label{eq:0-mom-eq}
\eeq
Using the identities specified in Eqs.~(\ref{eq:vectors})-(\ref{eq:thetas}), we can rewrite Eqs.~(\ref{eq:0-mom-eq}) as
\bqa
&&\frac{1}{1+\xi}\left( \cosh(\vartheta-\varsigma) \partial_\tau + \frac{\sinh(\vartheta-\varsigma)}{\tau} \partial_\varsigma \right) \xi
-\frac{6}{\phard}\left( \cosh(\vartheta-\varsigma) \partial_\tau + \frac{\sinh(\vartheta-\varsigma)}{\tau} \partial_\varsigma \right) \phard
\nonumber\\&&
\hspace{1cm}
-2 \left( \sinh(\vartheta-\varsigma) \partial_\tau + \frac{\cosh(\vartheta-\varsigma)}{\tau} \partial_\varsigma \right) \vartheta
= 2 \Gamma \left[ 1 - {\cal R}^{3/4}(\xi) \sqrt{1+\xi} \right] \, .
\label{eq:zerothmoment}
\eqa

\subsubsection{1st moment of the non-boost invariant Boltzmann equation}
\label{subsubsec:1st-Bolteqn}

In Sect.~\ref{subsec:moment} we pointed out the equivalence between the first moment of the Boltzmann equation and energy-momentum conservation (Eq.~(\ref{eq:ene-mom})). Therefore, we can use Eqs.~(\ref{eq:en-mom-eqs}) derived from the energy-momentum tensor given by~(\ref{Tmunudec}). 
\begin{subequations}
\label{eq:en-mom-eqs-2}
\begin{align}
\Bigl(\partial_\tau+\frac{\tanh (\vartheta-\varsigma)}{\tau}\,\partial_\varsigma \Bigr){\cal E}(\tau,\varsigma)&=\nn\\&\hspace{-2cm}-\bigl({\cal E}(\tau,\varsigma)+{\cal P}_L (\tau,\varsigma)\bigr)\Bigl(\tanh (\vartheta-\varsigma)\,\partial_\tau +\frac{\partial_\varsigma}{\tau} \Bigr)\vartheta\,,\\
\Bigl(\tanh (\vartheta-\varsigma) \partial_\tau+\frac{\partial_\varsigma}{\tau}\Bigr){\cal P}_L(\tau,\varsigma)&=\nn\\&\hspace{-2cm}-\bigl({\cal E}(\tau,\varsigma)+{\cal P}_L (\tau,\varsigma)\bigr)\Bigl(\partial_\tau+\frac{\tanh (\vartheta-\varsigma)}{\tau}\,\partial_\varsigma \Bigr)\vartheta\,.
\end{align}
\end{subequations}
Using the RS ansatz~(\ref{eq:rsansatz-long}) it is possible to obtain analytic expressions for the energy density and longitudinal pressure 
(see \ref{Ap:A}, Eqs.~(\ref{energyaniso}) and~(\ref{longpressaniso}), respectively). Using these results, plugging them into 
Eqs.~(\ref{eq:en-mom-eqs-2}) and performing some algebra, we finally obtain the remaining two evolution equations
\begin{subequations}
\label{eq:1stmoment}
\begin{align}
&\frac{{\cal R}'(\xi)}{{\cal R}(\xi)}\,\partial_\tau \xi\,+\,4\,\frac{\partial_\tau \phard}{\phard}\,+\,
\frac{\tanh (\vartheta-\varsigma)}{\tau}\,\biggl(\frac{{\cal R}'(\xi)}{{\cal R}(\xi)}\,\partial_\varsigma \xi+\,4\,
\frac{\partial_\varsigma \phard}{\phard}\biggr)\nn\\
&\hspace{4.5cm}=-\Bigl(1+\frac{1}{3}\frac{{\cal R}_L(\xi)}{{\cal R}(\xi)}\Bigr)\Bigl(\tanh (\vartheta-\varsigma)\,\partial_\tau+
\frac{\partial_\varsigma}{\tau}\Bigr)\vartheta\,,\label{eq:1stmoment-1}\\
&\tanh (\vartheta-\varsigma)\Biggl(\frac{{\cal R}'_L(\xi)}{{\cal R}_L(\xi)}\,\partial_\tau \xi\,+\,4\,
\frac{\partial_\tau \phard}{\phard}\Biggr)+\frac{1}{\tau}\biggl(\frac{{\cal R}'_L(\xi)}{{\cal R}_L(\xi)}\,\partial_\varsigma \xi\,+\,
4\,\frac{\partial_\varsigma \phard}{\phard}\biggr)
\nn\\
&\hspace{4.5cm}= -\Bigl(3\,\frac{{\cal R}(\xi)}{{\cal R}_L(\xi)}+1\Bigr)\Bigl(\partial_\tau+
\frac{\tanh (\vartheta-\varsigma)}{\tau}\,\partial_\varsigma \Bigr)\vartheta\,,\label{eq:1stmoment-2}
\end{align}
\end{subequations}
where ${\cal R}(\xi)$ and ${\cal R}_L(\xi)$ are defined in Eqs.~(\ref{energyaniso}) and~(\ref{longpressaniso}), respectively, while 
${\cal R}'(\xi)$ and ${\cal R}'_L(\xi)$ are 
\begin{eqnarray*}
{\cal R}'(\xi)&=&\partial_\xi {\cal R}(\xi)=\frac{1}{4}\biggl(\frac{1-\xi}{\xi(1+\xi)^2}-\frac{\text{arctan}
\sqrt{\xi}}{\xi^{3/2}}\biggr)\,,\\
{\cal R}'_L(\xi)&=& \frac{3}{4} \left( \frac{3 + 5 \xi}{\xi^2 (1 + \xi)^2} - \frac{3\arctan{\sqrt{\xi}}}{\xi^{5/2}} \right) .
\end{eqnarray*}
Eqs.~(\ref{eq:1stmoment}) together with (\ref{eq:zerothmoment}) give us three coupled partial differential equations for the 
parameters $\xi(\tau,\varsigma), \vartheta(\tau,\varsigma)$  and $\phard (\tau,\varsigma)$ of the RS ansatz~(\ref{eq:rsansatz-long}). 
Their evolution will describe the transition between early-time dynamics and late-time hydrodynamical behaviour for a non-boost-invariant 
plasma in a unified framework. 

The (0+1)-dimensional boost invariant case can be recovered within our approach if we require that $\vartheta(\tau,\varsigma)=\varsigma$, 
$\xi(\tau,\varsigma)=\xi(\tau)$ and $\phard(\tau,\varsigma)=\phard(\tau)$. By taking this limit in Eqs.~(\ref{eq:zerothmoment}) 
and~(\ref{eq:1stmoment}) and rearranging some terms, we 
obtain two differential equations for $\xi(\tau)$ and $\phard(\tau)$
\begin{subequations}
 \begin{align}
 \label{eq:bi1}
  \frac{1}{1+\xi} \partial_\tau \xi- \frac{2}{\tau} - \frac{6}{\phard} \partial_\tau \phard &= 
2 \Gamma \left[ 1 - {\cal R}^{3/4}(\xi) \sqrt{1+\xi} \right] \, ,\\
 \label{eq:bi2}
 \frac{{\cal R}'(\xi)}{{\cal R}(\xi)} \partial_\tau \xi + \frac{4}{\phard} \partial_\tau \phard &= 
\frac{1}{\tau} \left[ \frac{1}{\xi(1+\xi){\cal R}(\xi)} - \frac{1}{\xi} - 1 \right] \, .
 \end{align}
  \label{eq:bi}
\end{subequations}
The last expressions correspond precisely to the boost invariant case studied before by us (see Eqs.~(20) and~(22) in Ref.
~\cite{Martinez:2010sc}). For the boost invariant case, one can also show that free streaming and ideal hydrodynamical behaviour are
reproduced when $\Gamma=0$ and $\Gamma\to\infty$, respectively~\cite{Martinez:2010sc}. In addition, the (0+1)-dimensional 2nd-order
viscous hydrodynamics of Israel and Stewart
can be recovered from our ansatz in the small $\xi$ limit if we identify~\cite{Martinez:2010sc}\footnote{We refer the 
reader to Sect. 2.4 of Ref.~\cite{Martinez:2010sc} for further details.}
\begin{subequations}
\bqa
\Gamma &=& \frac{2}{\tau_\pi}\,, \\ 
\tau_\pi &=& \frac{5}{4}\frac{\eta}{\cal P}\,.
\eqa
\label{eq:hydromatch}
\end{subequations}
We will implement this matching in our numerical simulations of the non-boost-invariant evolution equations~(\ref{eq:zerothmoment}) 
and (\ref{eq:1stmoment}).  In this manner the evolution of the system is specified by the shear viscosity with no need to introduce
additional parameters, e.g. thermalization time scales, etc.


\begin{figure}[ht]
\begin{center}
\includegraphics[width=16cm]{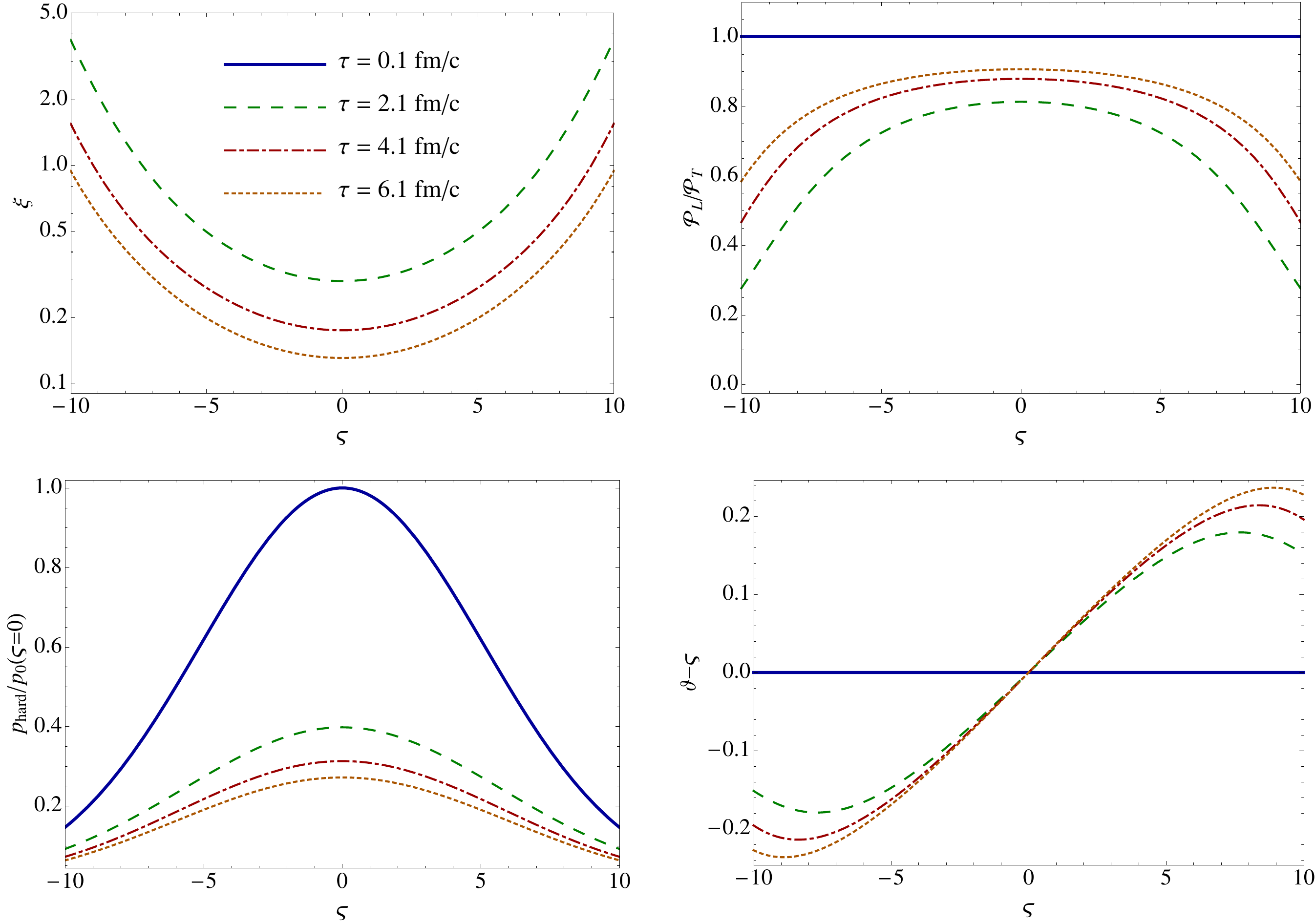}
\end{center}
\vspace{-6mm}
\caption{
Evolution of the rapidity dependence of the dynamical variables in the case of isotropic Gaussian initial conditions and strong coupling
viscosity corresponding to $\bar\eta = 1/(4\pi)$.
}
\label{fig:iso-gaussian-sc1}
\end{figure}

\begin{figure}[ht]
\begin{center}
\includegraphics[width=16cm]{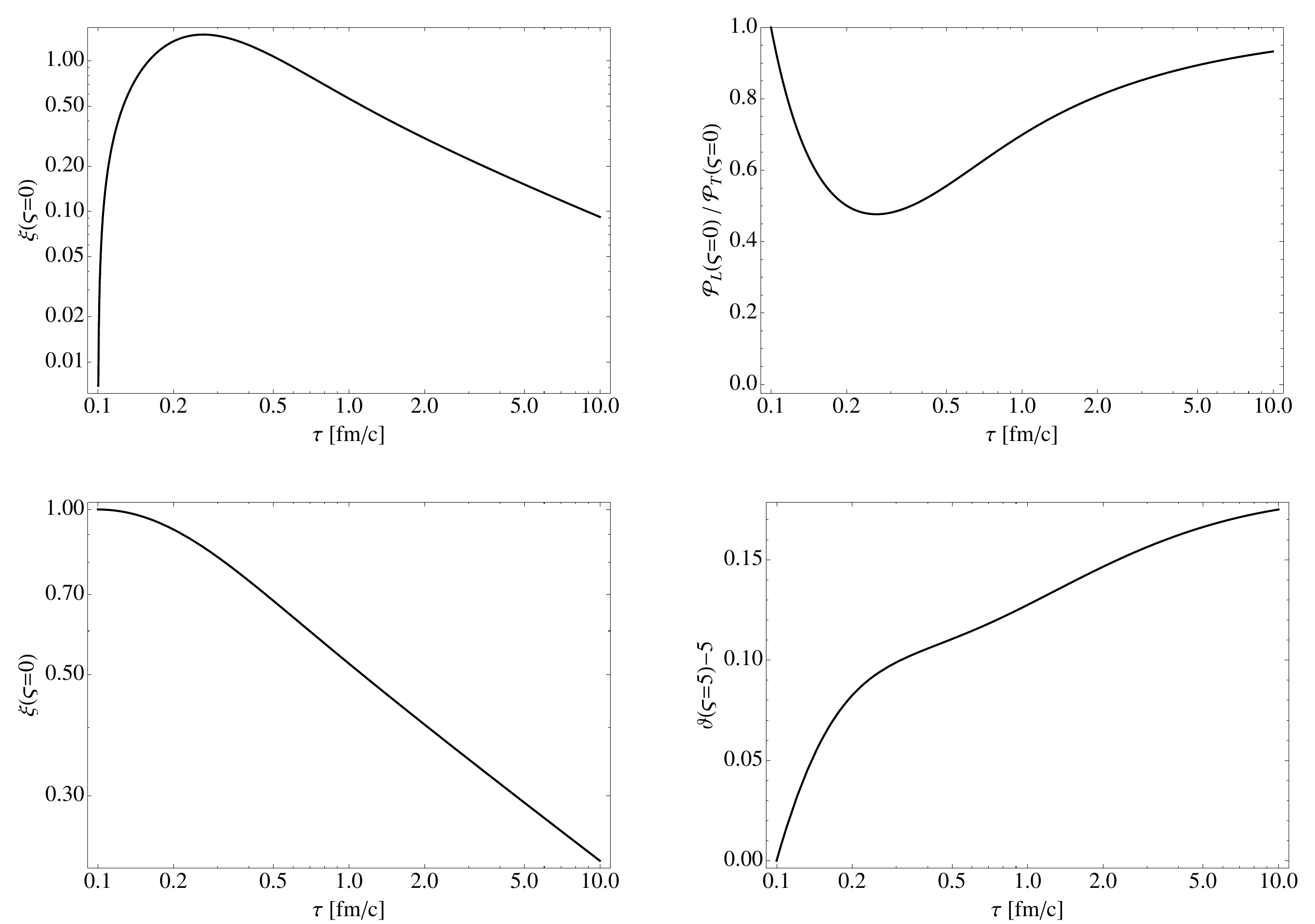}
\end{center}
\vspace{-6mm}
\caption{
Evolution of the dynamical variables in the case of isotropic Gaussian initial conditions and strong coupling
viscosity corresponding to $\bar\eta = 1/(4\pi)$.  All panels show evolution at central rapidity, except panel
showing time evolution of the hyperbolic angle $\theta-\varsigma$ which shows the evolution at $\varsigma=5$.
}
\label{fig:iso-gaussian-sc2}
\end{figure}

\begin{figure}[ht]
\begin{center}
\includegraphics[width=16cm]{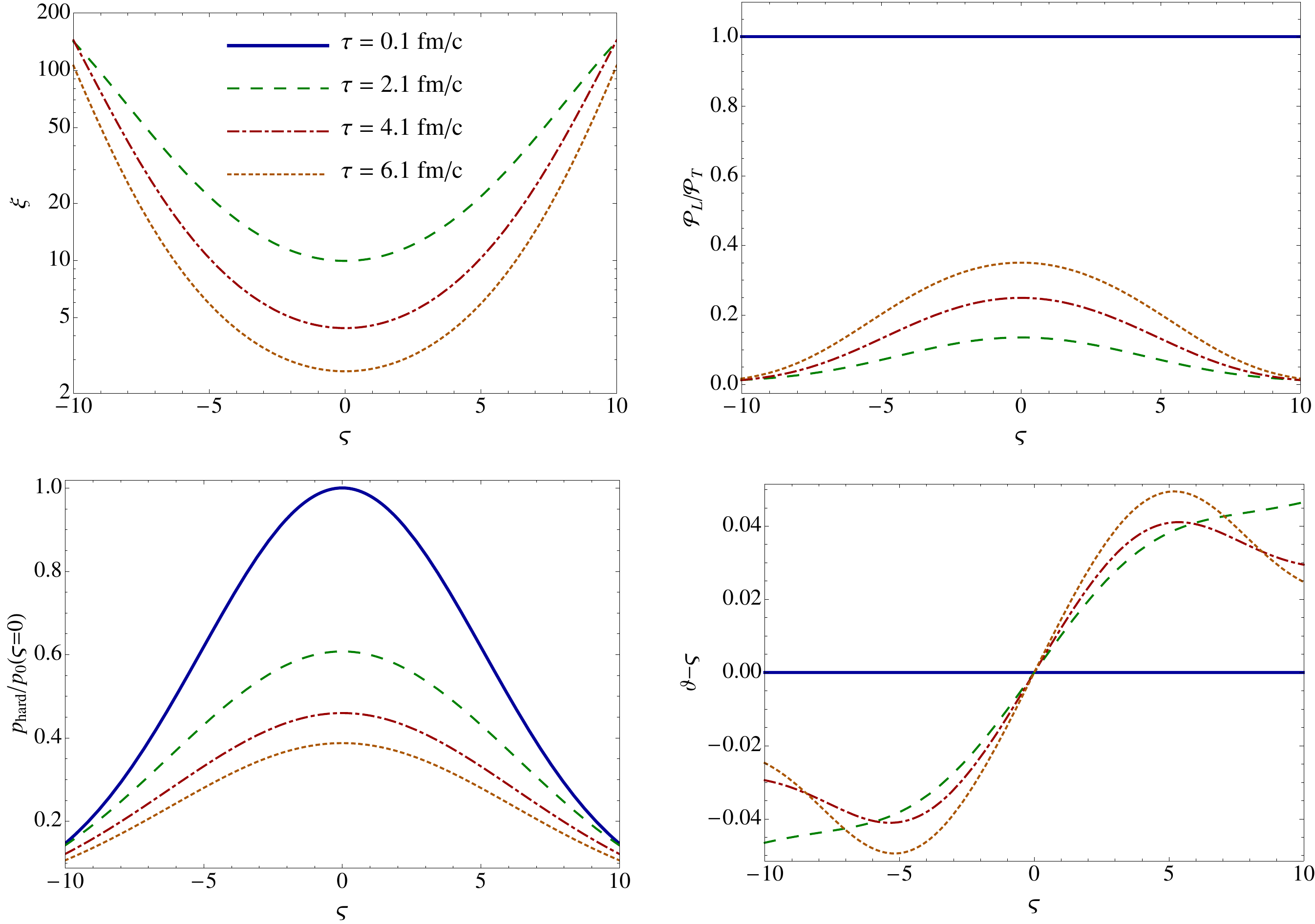}
\end{center}
\vspace{-6mm}
\caption{
Evolution of the rapidity dependence of the dynamical variables in the case of isotropic Gaussian initial conditions and weak coupling
viscosity corresponding to $\bar\eta = 10/(4\pi)$.
}
\label{fig:iso-gaussian-wc1}
\end{figure}

\begin{figure}[ht]
\begin{center}
\includegraphics[width=16cm]{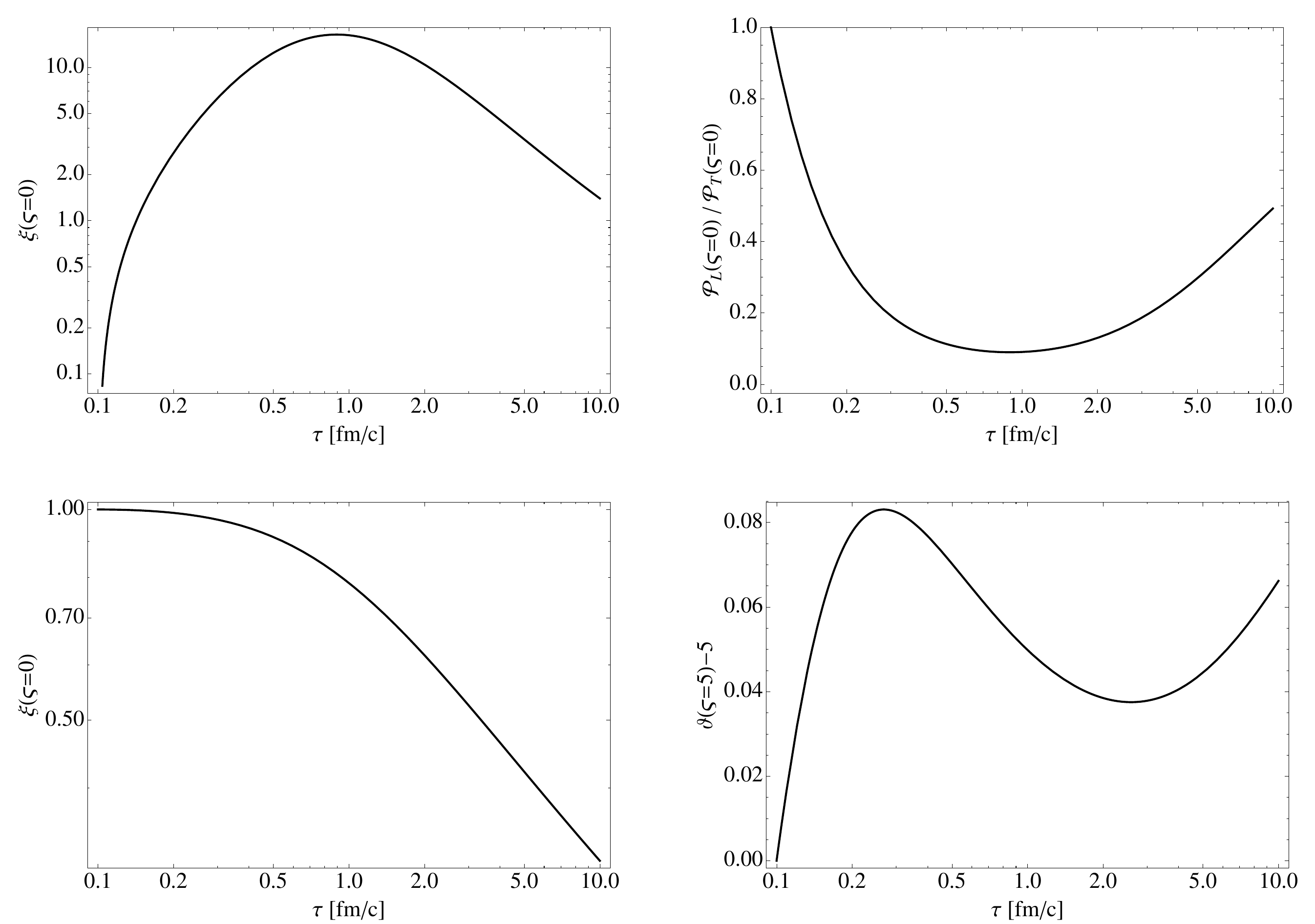}
\end{center}
\vspace{-6mm}
\caption{
Evolution of the dynamical variables in the case of isotropic Gaussian initial conditions and weak coupling
viscosity corresponding to $\bar\eta = 10/(4\pi)$.  All panels show evolution at central rapidity, except panel
showing time evolution of the hyperbolic angle $\theta-\varsigma$ which shows the evolution at $\varsigma=5$.
}
\label{fig:iso-gaussian-wc2}
\end{figure}

\begin{figure}[ht]
\begin{center}
\includegraphics[width=16cm]{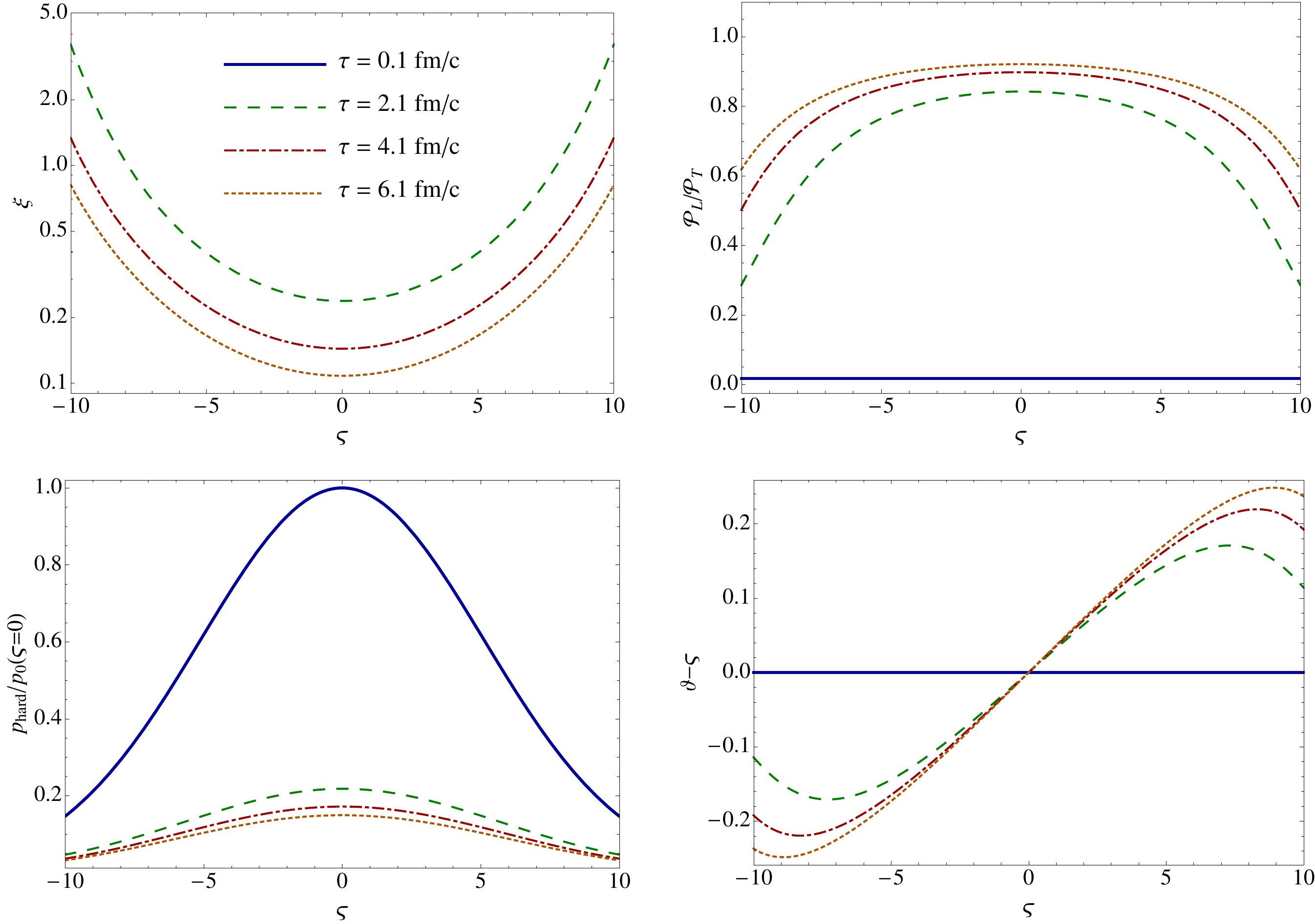}
\end{center}
\vspace{-6mm}
\caption{
Evolution of the rapidity dependence of the dynamical variables in the case of anisotropic Gaussian initial conditions and strong coupling
viscosity corresponding to $\bar\eta = 1/(4\pi)$.  Initial value of $\xi$ is $\xi_0 = 100$, independent of rapidity.
}
\label{fig:aniso-gaussian-sc1}
\end{figure}

\begin{figure}[ht]
\begin{center}
\includegraphics[width=16cm]{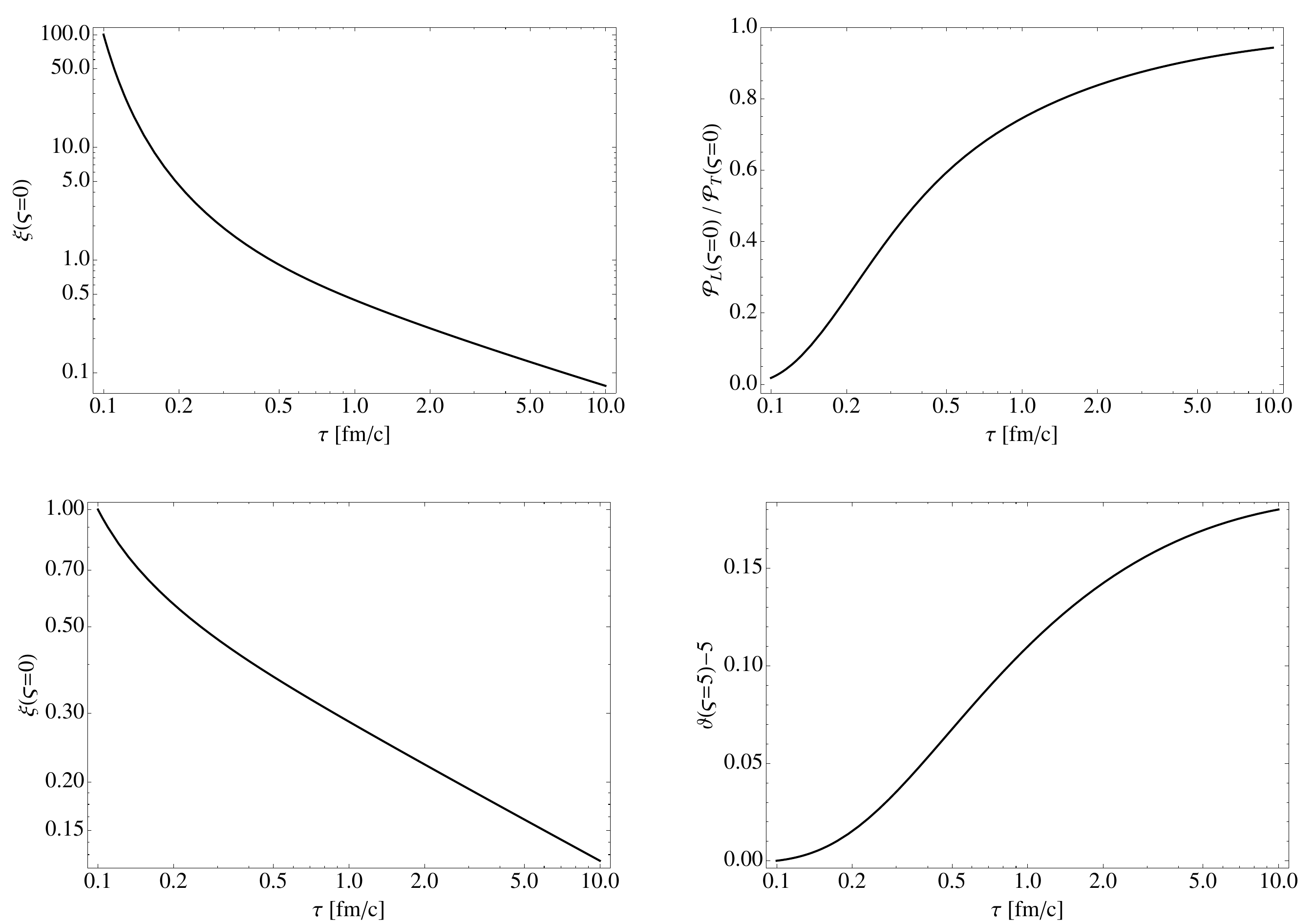}
\end{center}
\vspace{-6mm}
\caption{
Evolution of the dynamical variables in the case of anisotropic Gaussian initial conditions and strong coupling
viscosity corresponding to $\bar\eta = 1/(4\pi)$.  All panels show evolution at central rapidity, except panel
showing time evolution of the hyperbolic angle $\theta-\varsigma$ which shows the evolution at $\varsigma=5$.
Initial value of $\xi$ is $\xi_0 = 100$, independent of rapidity.
}
\label{fig:aniso-gaussian-sc2}
\end{figure}

\begin{figure}[ht]
\begin{center}
\includegraphics[width=16cm]{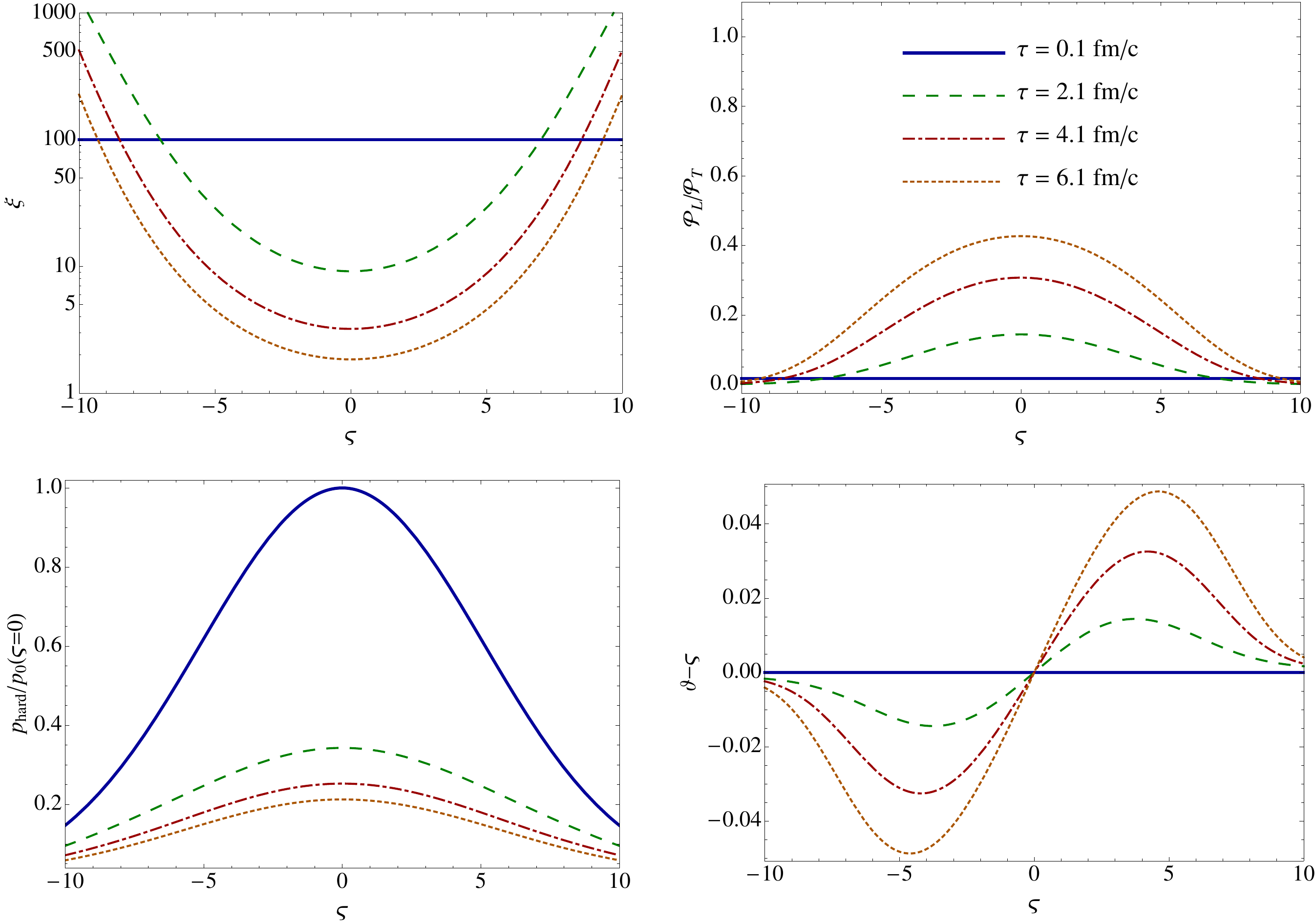}
\end{center}
\vspace{-6mm}
\caption{
Evolution of the rapidity dependence of the dynamical variables in the case of anisotropic Gaussian initial conditions and weak coupling
viscosity corresponding to $\bar\eta = 10/(4\pi)$.  Initial value of $\xi$ is $\xi_0 = 100$, independent of rapidity.
}
\label{fig:aniso-gaussian-wc1}
\end{figure}

\begin{figure}[ht]
\begin{center}
\includegraphics[width=16cm]{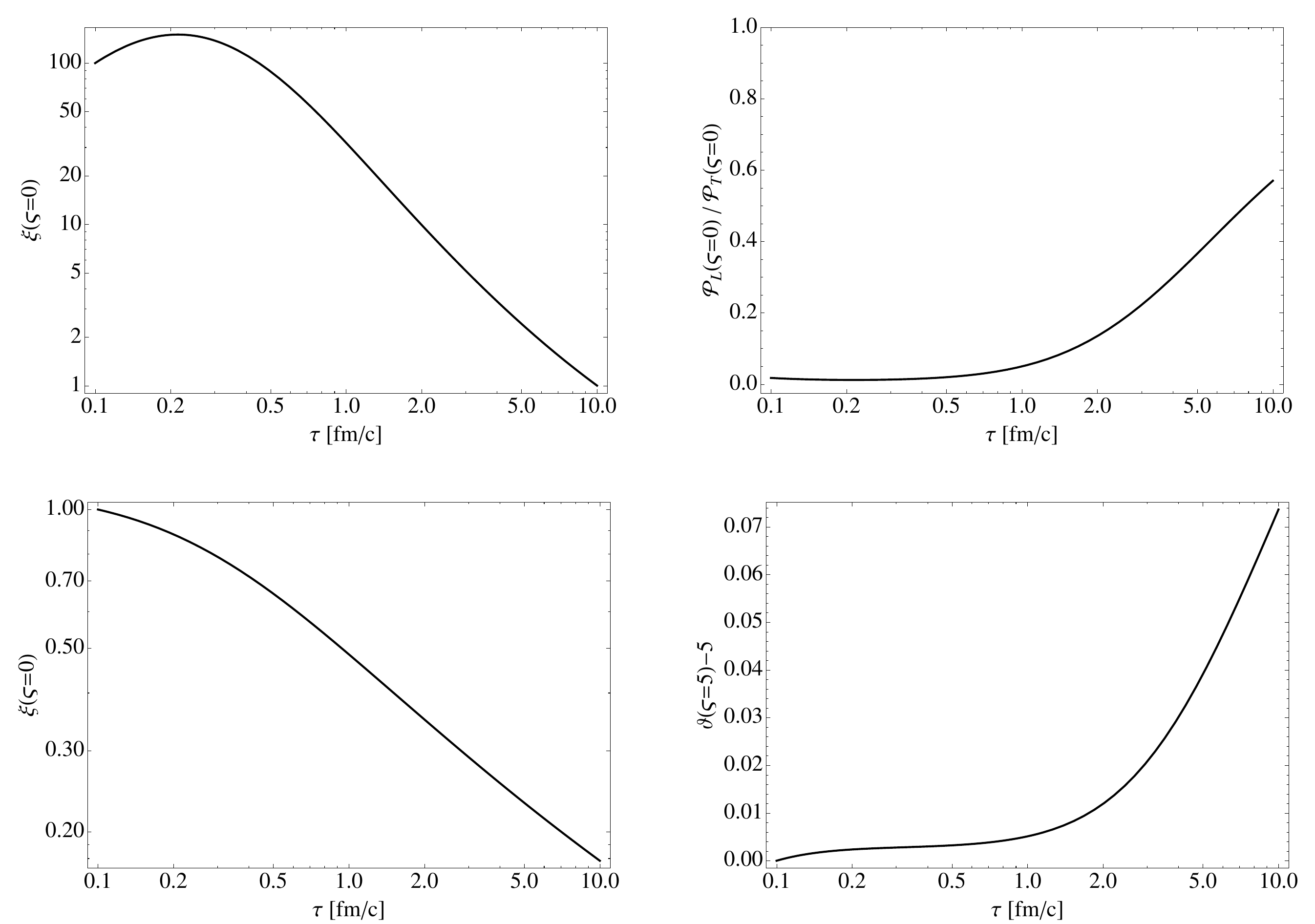}
\end{center}
\vspace{-6mm}
\caption{
Evolution of the dynamical variables in the case of anisotropic Gaussian initial conditions and weak coupling
viscosity corresponding to $\bar\eta = 10/(4\pi)$.  All panels show evolution at central rapidity, except panel
showing time evolution of the hyperbolic angle $\theta-\varsigma$ which shows the evolution at $\varsigma=5$.
Initial value of $\xi$ is $\xi_0 = 100$, independent of rapidity.
}
\label{fig:aniso-gaussian-wc2}
\end{figure}

\section{Results and Discussion}
\label{sec:results}

In this section we present the results of numerically integrating the coupled nonlinear differential equations 
using the relation between the relaxation rate $\Gamma$ and the ratio of shear viscosity to equilibrium entropy $\bar\eta \equiv \eta/{\cal S}$ which results from Eqs.~(\ref{eq:hydromatch}), namely
\beq
\Gamma = \frac{2T(\tau)}{5\bar\eta} = \frac{2{\cal R}^{1/4}(\xi)\phard}{5\bar\eta} \, ,
\label{eq:gammamatch}
\eeq
which should be used for $\Gamma$ in Eqs.~(\ref{eq:zerothmoment}) and (\ref{eq:1stmoment}).

\subsection{Initial Conditions}

In order to minimize the number of figures presented here we will only consider LHC-like initial conditions. Additionally,
for purposes of this paper we will present two types of initial conditions:  (I) an initially isotropic distribution which
has a Gaussian profile in the number density and (II) an initially anisotropic distribution which has the same Gaussian density profile
as case (I).
For the number density profile in spatial rapidity ($\varsigma$) we use a Gaussian which successfully describes experimentally observed pion 
rapidity spectra from AGS to RHIC energies \cite{Bearden:2004yx,Park:2001gm,Back:2005hs,Veres:2008nq,Bleicher:2005tb}
and extrapolate this result to LHC energies.  The parametrization we use is
\begin{equation}
\label{yprofile}
 n(\varsigma)= n_0 \exp \Biggl(-\frac{\varsigma^2}{2\sigma_\varsigma^2}\Biggr) \; ,
\end{equation}
with
\begin{equation}
\label{width}
 \sigma_\varsigma^2=\frac{8}{3}\frac{c_s^2}{(1-c_s^4)}\ln \left(\sqrt{s_{NN}}/2 m_p\right) \; , 
\end{equation}
where $c_s$ is the sound velocity, $m_p = 0.938$ GeV is the proton mass, for LHC $\sqrt{s_{NN}}= 5.5\;{\rm TeV}$ is the
nucleon-nucleon center-of-mass energy, and $n_0$ is the number density at central rapidity.
In this paper we will use an ideal equation of state for which $c_s = 1/\sqrt{3}$ in natural units.

In case (I) we can use Eq.~(\ref{yprofile}) to straightforwardly calculate the initial dependence of $\phard$
on rapidity by using Eq.~(\ref{nanisio}) with $\xi(\varsigma,\tau=\tau_0)=0$ to obtain
\bqa 
\phard(\varsigma,\tau=0) = p_0 \left[ \exp \Biggl(-\frac{\varsigma^2}{2\sigma_\varsigma^2}\Biggr) \right]^{1/3} \; ,
\eqa
where $p_0 =  \phard(\varsigma=0,\tau=0)$ is the initial ``temperature'' at central rapidity.  For LHC conditions
we use $p_0 = 845$ MeV at an initial time of $\tau_0 = $ 0.1 fm/c.
In case (II) we choose $\xi(\varsigma,\tau=\tau_0) = 100$ which means that we should increase the central value of
$\phard$ by a factor of $(1+\xi)^{1/6} \simeq 2.16$ in order to start with the same initial particle density distribution
[see Eq.~(\ref{nanisio})].  Doing this
gives in case (II) $p_0 \simeq 1.82$ GeV.

In both cases (I) and (II) we must also fix the initial condition for the hyperbolic angle $\vartheta$
which specifies the four-velocity of the local rest frame.  Here we choose $\vartheta(\varsigma,\tau=\tau_0)=\varsigma$
following Refs.~\cite{Ryblewski:2010bs,Satarov:2006jq,Satarov:2006iw}.  
This is not the most general initial condition possible, of course; however, absent some 
guiding principle we choose here to follow Refs.~\cite{Ryblewski:2010bs,Satarov:2006jq,Satarov:2006iw} and approximate the initial state
as having a four-velocity profile corresponding to an initially boost-invariant configuration.

\subsection{Case I:  Isotropic Initial Conditions}

In this subsection we present results for the spatial-rapidity and proper-time dependence of our dynamic parameters 
in the case that the parton distribution function is assumed to be locally isotropic at $\tau = \tau_0 = 0.1$ fm/c.
We present the cases of transport coefficients corresponding to a strongly-coupled
plasma and a weakly-coupled plasma separately.

\subsubsection{Strongly Coupled}

In Fig.~\ref{fig:iso-gaussian-sc1} we plot the spatial-rapidity dependence of the anisotropy parameter (upper left), 
the ratio of the longitudinal to transverse pressure (upper right), the hard momentum scale (lower left), and
the hyperbolic angle of the local rest frame four-velocity as measured in the lab frame (lower right).  We have
fixed the ratio of the shear viscosity to entropy density to be $\bar\eta \equiv \eta/{\cal S} = 1/(4\pi)$.  Each
line in the plot corresponds to a fixed proper-time $\tau \in \{ 0.1,\,2.1,\,4.1,\,6.1 \}$ fm/c.    From these
figures we see that large momentum-space anisotropies are developed at large rapidity.  This can be traced
back to the fact that the initial temperature is lower at large rapidity, causing the rate of relaxation back to
isotropy to be slower \cite{Martinez:2009mf}.  At $\tau = 6.1$ fm/c we see that in the case of strong-coupling
shear viscosity at $\varsigma=5$, ${\cal P}_L/{\cal P}_T \simeq 0.82$.  

Some readers may
be confused by the fact that, although we started from an isotropic initial condition, the system did not remain
isotropic at all times.  The deviation from isotropy is due to the dynamical expansion of the system resulting 
from the initial fluid longitudinal velocity profile.  This 
can be seen most clearly by examining the boost-invariant limits presented in Eqs.~(\ref{eq:bi}).
While it is true that when $\xi=0$ the right hand side of Eq.~(\ref{eq:bi1}) vanishes, there are terms on the left
hand side which drive $\xi$ away from zero.  In the case of ideal hydrodynamical expansion, one would have
$p_{\rm hard} \propto \tau^{-1/3}$ and the second and third terms on the left hand side of Eq.~(\ref{eq:bi1})
would cancel with one another.  However, when non-ideal corrections are taken into account, one finds
a different scaling coefficient for the hard momentum scale.  For example, assuming $p_{\rm hard} \propto \tau^{-\alpha}$
the second and third terms would combine to give a term proportional to $(3\alpha-1)/\tau$.  For any $\alpha<1/3$
this term is negative and therefore represents a source for increasing the anisotropy parameter in time.  The 
same qualitative effect can be seen in solutions to second-order viscous hydrodynamics.  A detailed comparison
of the evolution of the pressure anisotropy using our formalism and second-order viscous hydrodynamics is
presented in Ref.~\cite{Martinez:2010sc} (see, in particular, Figs.~4 and 5 where isotropic initial conditions
are also assumed).  We also reemphasize that in 
the limit of small anisotropy the dynamical equations presented here reduce identically to second order
viscous hydrodynamics.  For the proof we refer the reader to Sec.~2.6 of Ref.~\cite{Martinez:2010sc}.

We also note that the
deviation of $\vartheta \neq \varsigma$ which results for $\tau > \tau_0$ is indicative of the breaking
of longitudinal boost invariance.  The fact that $\vartheta > \varsigma$ results from there being pressure
gradients in the longitudinal direction which increase the longitudinal velocity beyond that which would
be obtained if the flow were boost invariant. The fact that the $\vartheta-\varsigma$ curves turn over at large
spatial rapidity is an indication that the material at the edges is propagating at a slower longitudinal velocity
than the material in the central region.  We also note that as shown in the lower left
plot of the hard momentum scale both the width and amplitude of the momentum distribution are changing
with proper-time.  The is width increasing with proper-time due to longitudinal
pressure gradients.

In Fig.~\ref{fig:iso-gaussian-sc2} we plot the proper-time dependence of our dynamical variables and
the pressure anisotropy for fixed spatial rapidity.  We show the anisotropy parameter at central rapidity
(upper left), the ratio of the longitudinal to transverse pressure at central rapidity (upper right), the hard
momentum scale at central rapidity  (lower left), and the hyperbolic angle of the local rest frame four-velocity
as measured in the lab frame at $\varsigma=5$ (lower right).  We do not show $\vartheta$ at central rapidity
since $\vartheta(\varsigma=0)=0$ by symmetry \cite{Satarov:2006jq,Satarov:2006iw}.  As one can see
from the plots presented in Fig.~\ref{fig:iso-gaussian-sc2} at central rapidities large momentum-space
anisotropies can develop at early times even in the case of a strongly-coupled plasma.  From the upper right
panel of Fig.~\ref{fig:iso-gaussian-sc2} we see that at $\tau \simeq 0.25$ fm/c one has ${\cal P}_L/{\cal P}_T \lsim 0.5$.  
Such large pressure anisotropies could have a significant effect on observables which
are sensitive to the early-time dynamics of the quark-gluon plasma such as photon 
\cite{Schenke:2006yp,Ipp:2007ng,Bhattacharya:2008mv,Dusling:2009bc,Ipp:2009ja,Bhattacharya:2009sb,Bhattacharya:2010sq},
dilepton \cite{Mauricio:2007vz,Martinez:2008di,Martinez:2008mc,Dusling:2008xj} and heavy-quarkonium production 
\cite{Dumitru:2007hy,Dumitru:2009ni,Dumitru:2009fy,Baier:2008js,%
Noronha:2009ia,Burnier:2009yu,Carrington:2009vm,Philipsen:2009wg,Chandra:2010xg,Roy:2010zg}.

\subsubsection{Weakly Coupled}

In Fig.~\ref{fig:iso-gaussian-wc1} we plot the spatial-rapidity dependence of the anisotropy parameter (upper left), 
the ratio of the longitudinal to transverse pressure (upper right), the hard momentum scale (lower left), and
the hyperbolic angle of the local rest frame four-velocity as measured in the lab frame (lower right).  We have
fixed the ratio of the shear viscosity to entropy density to be $\bar\eta \equiv \eta/{\cal S} = 10/(4\pi)$.  Each
line in the plot corresponds to a fixed proper-time $\tau \in \{ 0.1,\,2.1,\,4.1,\,6.1 \}$ fm/c.    From these
figures we see that large momentum-space anisotropies are developed at all rapidities.  At $\tau = 6.1$ fm/c we
see that in the case of weak-coupling shear viscosity ${\cal P}_L/{\cal P}_T \lsim 0.4$ at all rapidities.  In
terms of the hyperbolic angle $\vartheta$ we see weaker deviations from boost-invariant longitudinal flow
than seen in the case of strong-coupling transport coefficients (Fig.~\ref{fig:iso-gaussian-sc1}).  Such strong
momentum-space anisotropies indicate that a naive viscous hydrodynamical treatment would not be trustworthy at
any spatial rapidity shown.  This is in line with the results presented in Ref.~\cite{Martinez:2009mf} where
we discussed the generation of negative longitudinal pressure due to large momentum-space anisotropies
when using 2nd-order viscous hydrodynamical evolution equations.  Using the partial-differential equations
presented here we are able to evolve the system even in the case of large momentum-space anisotropies
and the longitudinal pressure is guaranteed to be positive at all times.

In Fig.~\ref{fig:iso-gaussian-wc2} we plot the proper-time dependence of our dynamical variables and
the pressure anisotropy for fixed spatial rapidity.  We show the anisotropy parameter at central rapidity
(upper left), the ratio of the longitudinal to transverse pressure at central rapidity (upper right), the hard
momentum scale at central rapidity  (lower left), and the hyperbolic angle of the local rest frame four-velocity
as measured in the lab frame at $\varsigma=5$ (lower right).  As one can see
from the plots presented in Fig.~\ref{fig:iso-gaussian-wc2} at central rapidities large momentum-space
anisotropies persist at all times.  From the upper right
panel of Fig.~\ref{fig:iso-gaussian-wc2} we see that at $\tau \simeq 0.25$ fm/c one has ${\cal P}_L/{\cal P}_T \lsim 0.25$.
This is much larger than the strongly-coupled case presented in the previous section.  This suggests 
that anisotropic photon, dilepton, and heavy-quarkonium production rates integrated in spacetime using 
our dynamical model could be used to constrain quark-gluon transport coefficients.  This could be done by 
comparing theoretical predictions resulting from the assumption of strong-coupling (Fig.~\ref{fig:iso-gaussian-sc1})
and weak-coupling (Fig.~\ref{fig:iso-gaussian-wc1}) transport coefficients and integrating over the resulting
spacetime evolution.  Finally we point out that one sees from the lower left panel of Fig.~\ref{fig:iso-gaussian-wc2} 
that the evolution of the hard momentum scale is far from that expected due to weakly-viscous expansion.
Initially we see a period of approximate free-streaming expansion which later transitions to viscous 
hydrodynamical expansion.  This would have the effect of increasing the lifetime of the quark-gluon
plasma in the weakly-coupled case and could result in stronger transverse flow than naively predicted
by a 2nd-order viscous hydrodynamical treatment.

\subsection{Case II:  Anisotropic Initial Conditions}

In this subsection we present results for the spatial-rapidity and proper-time dependence of our dynamical parameters 
in the case that the parton distribution function is assumed to be locally anisotropic at $\tau = \tau_0 = 0.1$ fm/c.
As detailed in the beginning of the results section we take $\xi(\tau_0) = 100$ which corresponds to a highly
oblate distribution with ${\cal P}_L \ll {\cal P}_T$.  
Such highly oblate distributions are predicted by color-glass-condensate models of the early-time dynamics of
a heavy-ion collision \cite{Baier:2000sb,Gelis:2010nm}.
As in the previous section that assumed isotropic initial conditions, we present
the cases of transport coefficients corresponding to a strongly-coupled plasma and a weakly-coupled plasma separately.

\subsubsection{Strongly Coupled}

In Fig.~\ref{fig:aniso-gaussian-sc1} we plot the spatial-rapidity dependence of the anisotropy parameter (upper left), 
the ratio of the longitudinal to transverse pressure (upper right), the hard momentum scale (lower left), and
the hyperbolic angle of the local rest frame four-velocity as measured in the lab frame (lower right).  We have
fixed the ratio of the shear viscosity to entropy density to be $\bar\eta \equiv \eta/{\cal S} = 1/(4\pi)$.  Each
line in the plot corresponds to a fixed proper-time $\tau \in \{ 0.1,\,2.1,\,4.1,\,6.1 \}$ fm/c.    From these
figures we see that large momentum-space anisotropies persist at large rapidity.  At $\tau = 6.1$ fm/c 
we see that in the case of strong-coupling shear viscosity at $\varsigma=5$, ${\cal P}_L/{\cal P}_T \simeq 0.85$.
This is similar in magnitude to the result obtained when one starts with an isotropic initial condition 
(see Fig.~\ref{fig:iso-gaussian-sc1}).  We also note a characteristic triangular shape to the spatial-rapidity
dependence of the pressure anisotropy (Fig.~\ref{fig:aniso-gaussian-sc1} top right).

In Fig.~\ref{fig:aniso-gaussian-sc2} we plot the proper-time dependence of our dynamical variables and
the pressure anisotropy for fixed spatial-rapidity.  We show the anisotropy parameter at central rapidity
(upper left), the ratio of the longitudinal to transverse pressure at central rapidity (upper right), the hard
momentum scale at central rapidity  (lower left), and the hyperbolic angle of the local rest frame four-velocity
as measured in the lab frame at $\varsigma=$ (lower right).  As one can see
from the plots presented in Fig.~\ref{fig:aniso-gaussian-sc2} at central rapidities large momentum-space
anisotropies persist in the case of a strongly-coupled plasma.  From the upper right
panel of Fig.~\ref{fig:aniso-gaussian-sc2} we see that at $\tau \simeq 0.25$ fm/c one has ${\cal P}_L/{\cal P}_T
\lsim 0.45$.  This is similar to the case of isotropic initial conditions suggesting a kind of ``attractor'' for the
momentum-space anisotropy evolution.  This attractor can be identified as the Navier-Stokes solution if
one analyzes the late time asymptotic solutions of the evolution equations presented here and/or 
2nd-order viscous hydrodynamics.

\subsubsection{Weakly Coupled}

In Fig.~\ref{fig:aniso-gaussian-wc1} we plot the spatial-rapidity dependence of the anisotropy parameter (upper left), 
the ratio of the longitudinal to transverse pressure (upper right), the hard momentum scale (lower left), and
the hyperbolic angle of the local rest frame four-velocity as measured in the lab frame (lower right).  We have
fixed the ratio of the shear viscosity to entropy density to be $\bar\eta \equiv \eta/{\cal S} = 10/(4\pi)$.  Each
line in the plot corresponds to a fixed proper-time $\tau \in \{ 0.1,\,2.1,\,4.1,\,6.1 \}$ fm/c.    From these
figures we see that large momentum-space anisotropies are developed at all rapidities.  At $\tau = 6.1$ fm/c we
see that in the case of weak-coupling shear viscosity at $\varsigma=5$, ${\cal P}_L/{\cal P}_T \simeq 0.2$.

In Fig.~\ref{fig:aniso-gaussian-wc2} we plot the proper-time dependence of our dynamical variables and
the pressure anisotropy for fixed spatial rapidity.  We show the anisotropy parameter at central rapidity
(upper left), the ratio of the longitudinal to transverse pressure at central rapidity (upper right), the hard
momentum scale at central rapidity  (lower left), and the hyperbolic angle of the local rest frame four-velocity
as measured in the lab frame at $\varsigma=$ (lower right).  As one can see
from the plots presented in Fig.~\ref{fig:aniso-gaussian-wc2} at central rapidities large momentum-space
anisotropies persist for a long time in the case of an initially anisotropic weakly-coupled quark-gluon plasma.  
From the upper right panel of Fig.~\ref{fig:aniso-gaussian-wc2} we see that at $\tau \simeq 0.25$ fm/c one has 
${\cal P}_L/{\cal P}_T \lsim 0.01$.  Such extreme momentum-space anisotropies would preclude the use of a
viscous hydrodynamical treatment; however, our reorganized expansion offers some hope to treat systems with such
high anisotropies.


\section{Conclusions and Outlook}
\label{sec:conclusion}

In this paper we have derived three coupled partial differential equations given in Eqs.~(\ref{eq:zerothmoment}) 
and (\ref{eq:1stmoment}) whose solution gives the time evolution of the plasma anisotropy parameter $\xi$,
the typical hard momentum scale of the partons $\phard$, and the longitudinal flow velocity variable $\vartheta$.
We relaxed the assumption of boost-invariance by allowing the longitudinal flow velocity variable to differ from
the spatial rapidity $\varsigma$ and we also allowed the hard momentum scale and plasma anisotropy to depend on
the spatial rapidity.  The partial differential equations were obtained by taking moments of the Boltzmann
equation using a relaxation time approximation collisional kernel and a spheroidally anisotropic ansatz for the 
underlying partonic distribution function.  By requiring that our equations reduced
to 2nd order viscous hydrodynamics in the limit of small anisotropy we
were able to obtain an analytic connection between the relaxation rate $\Gamma$ appearing in the 
collisional kernel and the plasma shear viscosity $\eta$ and shear relaxation time $\tau_\pi$.  Using
Eqs.~(\ref{eq:zerothmoment}) and (\ref{eq:1stmoment}) and $\Gamma$ as a functions of $\eta$ we were able to 
numerically solve the coupled partial differential equations in both the strong and weak coupling limits.  

Our numerical results indicate that in both the strongly and weakly coupled 
cases large momentum-space anisotropies can be generated.  Therefore,
a treatment such as the one derived here in which momentum-space anisotropies are built into the leading order
ansatz can better describe the time evolution of the system.  Relatedly we demonstrated that within our reorganized
approach at all spatial rapidity the
longitudinal pressure remains positive during the entire evolution.  This is to be contrasted to 2nd-order viscous
hydrodynamics which can give negative pressures indicating the breakdown of the expansion about an 
isotropic equilibrium state.  We have also discussed the fact that our equations can reproduce both extreme limits of
the dynamics:  ideal hydrodynamical expansion when the shear viscosity goes to zero and free streaming when the
shear viscosity goes to infinity.  

The numerical results obtained here 
differ from those obtained in Ref.~\cite{Ryblewski:2010bs} in that we see a much slower relaxation
to isotropy at large spatial rapidity.  This results from the fact that the authors of Ref.~\cite{Ryblewski:2010bs}
assumed that there is a single rapidity-independent thermalization time scale which they arbitrarily chose to be 0.25 
fm/c.  In this work we instead implemented a method for matching to 2nd-order viscous 
hydrodynamics that was introduced by us in Ref.~\cite{Martinez:2010sc}.  The conclusion of that paper was that
the relaxation rate $\Gamma$ should be proportional to hard momentum scale (see Eq.~(\ref{eq:gammamatch}) herein).  
If in a certain region the average momentum scale is lower, as is the case in the forward regions described
by large spatial rapidity, this implies a slower relaxation to isotropic equilibrium.  In our case this 
naturally arises due to the matching to 2nd-order viscous hydrodynamics in the limit of small anisotropy
and as a result it not necessary to introduce external parameters such as thermalization times as a function of rapidity.

Beside the interesting theoretical goal of improving the description of systems which have large momentum-space
anisotropies, we point out that the dynamical equations derived here can be used to assess the impact of rapidity-dependent
momentum-space anisotropies on  observables which are sensitive to the early-time dynamics of the quark-gluon plasma such as photon 
\cite{Schenke:2006yp,Ipp:2007ng,Bhattacharya:2008mv,Dusling:2009bc,Ipp:2009ja,Bhattacharya:2009sb,Bhattacharya:2010sq}, 
dilepton \cite{Mauricio:2007vz,Martinez:2008di,Martinez:2008mc,Dusling:2008xj} and heavy-quarkonium production 
\cite{Dumitru:2007hy,Dumitru:2009ni,Dumitru:2009fy,Baier:2008js,%
Noronha:2009ia,Burnier:2009yu,Carrington:2009vm,Philipsen:2009wg,Chandra:2010xg,Roy:2010zg}.
Additionally, one could consider using the evolution of $\xi$ and $\phard$ presented here in order to fold in a 
realistic anisotropy evolution into simulations of non-abelian plasma instabilities \cite{Mrowczynski1993118,%
Romatschke:2003ms,Arnold:2003rq,Rebhan:2004ur,Arnold:2005vb,Rebhan:2005re,Bodeker:2007fw,Berges:2007re,Rebhan:2008uj}.

Looking forward it is of critical importance to include the transverse expansion and elliptical flow 
of the system.  This can be done minimally by allowing for one more momentum-space anisotropy parameter
that describes the momentum-space anisotropy in the transverse plane and allowing all dynamical parameters
to depend on proper-time, spatial rapidity, and transverse position.  Work along these lines is currently
underway \cite{Mart-Strick}.


\section*{Acknowledgments}
We thank R.~Ryblewski for reporting an error in the published version of this manuscript and J.~Himmelsbach for pointing out a typo in the expression for ${\cal R}_L^\prime$.
Additionally, we thank G.~Denicol, A.~Dumitru, and I.~Mishustin for useful discussions.  We thank D.~Bazow for checking
our numerical results.
M. Martinez and M. Strickland were supported by the Helmholtz International Center for FAIR 
Landesoffensive zur Entwicklung Wissenschaftlich-\"Okonomischer Exzellenz program.

\appendix

\section{Analytic expressions for the components of the energy momentum tensor, number density, and entropy density}
\label{Ap:A}
The energy-momentum tensor $T^{\mu\nu}= (2\pi)^{-3}\,\int d^3{\bf p}/p^0\, p^\mu p^\nu f({x},{p},t)$ for the RS ansatz
~(\ref{eq:rsansatz-long}) is diagonal in the comoving frame and its components are~\cite{Martinez:2010sc,Martinez:2009ry}
\begin{subequations}
\label{momentsanisotropic}
\begin{align}
\label{energyaniso}
{\cal E}(p_{\rm hard},\xi) &= T^{\tau\tau} \;= \frac{1}{2}\left(\frac{1}{1+\xi}
+\frac{\arctan\sqrt{\xi}}{\sqrt{\xi}} \right) {\cal E}_{\rm iso}(p_{\rm hard}) \; , \\ \nonumber
&\equiv{\cal R}(\xi)\,{\cal E}_{\rm iso}(p_{\rm hard})\, ,\\
\label{transpressaniso}
{\cal P}_T(p_{\rm hard},\xi) &= \frac{1}{2}\left( T^{xx} + T^{yy}\right) 
= \frac{3}{2 \xi} 
\left( \frac{1+(\xi^2-1){\cal R}(\xi)}{\xi + 1}\right)
 {\cal P}_T^{\rm iso}(p_{\rm hard}) \, , 
\\ \nonumber
&\equiv{\cal R}_{\rm T}(\xi){\cal P}_T^{\rm iso}(p_{\rm hard})\, , \\
\label{longpressaniso}
{\cal P}_L(p_{\rm hard},\xi) &= - T^{\varsigma}_\varsigma= \frac{3}{\xi} 
\left( \frac{(\xi+1){\cal R}(\xi)-1}{\xi+1}\right) {\cal P}_L^{\rm iso}(p_{\rm hard}) \; ,\\ \nonumber
&\equiv {\cal R}_{\rm L}(\xi){\cal P}_L^{\rm iso}(p_{\rm hard})\, ,
\end{align}
\end{subequations}
where ${\cal P}_T^{\rm iso}(p_{\rm hard})$ and ${\cal P}_L^{\rm iso}(p_{\rm hard})$ are the isotropic transverse and longitudinal 
pressures and ${\cal E}_{\rm iso}(p_{\rm hard})$ is the isotropic energy density.

One can also easily obtain the number density as a function of $p_{\rm hard}$ and $\xi$
\beq
n(\xi,\phard) = \frac{n_{\rm iso}(\phard)}{\sqrt{1+\xi}} \propto \frac{\phard^3}{\sqrt{1+\xi}} \, ,
\label{nanisio}
\eeq
where $n_{\rm iso}$ is the isotropic number density.  Similarly, one can obtain for the entropy density
\beq
{\cal S}(\xi,\phard) = \frac{{\cal S}_{\rm iso}(\phard)}{\sqrt{1+\xi}} \, .
\label{sanisio}
\eeq

\bibliographystyle{utphys}
\bibliography{strickland}

\end{document}